\documentclass[twocolumn,showpacs,preprintnumbers,amsmath,amssymb]{revtex4}

\usepackage{graphicx}
\usepackage{dcolumn}
\usepackage{bm}

\begin{document}

\title{Onset of classical
behaviour after a phase transition}

\author{R.J. Rivers$^{1}$}
\email{r.rivers@imperial.ac.uk}
\author{F.C. Lombardo$^{2}$}%
\email{lombardo@df.uba.ar}

\affiliation{$^{1}$Theoretical Physics Group; Blackett Laboratory,
Imperial College, London SW7 2BZ} \affiliation{$^2$ Departamento
de F\'\i sica, Facultad de Ciencias Exactas y Naturales,
Universidad de Buenos Aires - Ciudad Universitaria, Pabell\' on I,
1428 Buenos Aires, Argentina}
\date{\today}



\begin{abstract}

We analyze the onset of classical behaviour in a scalar field
after a continuous phase transition, in which the system-field,
the long wavelength order parameter of the model, interacts with
an environment, of its own short-wavelength modes. We compute the
decoherence time for the system-field modes from the master
equation and compare it with the other time scales of the model.
Within our approximations the decoherence time is in general the
smallest dynamical time scale. Demanding diagonalisation of the
decoherence functional produces identical results. The inclusion
of other environmental fields makes diagonalisation occur even
earlier.

\end{abstract}
\maketitle
03.70.+k,  05.70.Fh, 03.65.Yz

\section{Introduction}
The standard big bang cosmological model of the early universe
assumes a period of rapid cooling, giving a strong likelihood of
phase transitions, at the grand unified and electroweak
scales \cite{Kolb} in particular. What interests us in this talk is
the way in which phase transitions naturally take us from a
quantum to classical description of the universe.

That (continuous) transitions should move us rapidly to classical
behaviour is not surprising. Classical behaviour has two
attributes:
 \begin{itemize}
 \item
Classical correlations: By this is meant that the Wigner
function(al) $W[\pi ,\phi]$ peaks on classical phase-space
trajectories, with a probabilistic interpretation.
 \item
 Diagonalisation: By this is meant that the density matrix $\rho
 (t)$ should become (approximately) diagonal, in this case in a
 field basis. Alternatively, we can demand diagonalisation of the
 decoherence functional. In either case a probabilistic description (no
 quantum interference)
  is obtained.
 \end{itemize}

\noindent From the papers of Guth and Pi \cite{guthpi} onwards, it
has been appreciated that unstable modes lead to correlations
through squeezing. On the other hand, we understand
diagonalisation to be an almost inevitable consequence of tracing
over the 'environment' of the 'system' modes.

Continuous transitions supply both ingredients. Firstly, the field
ordering after such a transition is due to the growth in amplitude
of unstable long-wavelength modes, which arise automatically from
unstable maxima in the potential. Secondly, the stable
short-wavelength modes of the field, together with all the other
fields with which it interacts, form an environment whose
coarse-graining enforces diagonalisation and makes the
long-wavelength modes decohere.

While there are few doubts about the classical outcome, to
quantify these general observations is difficult because, with
fields, we are dealing with infinite degree of freedom systems.
One of us (F.L) has shown elsewhere \cite{diana} how classical
correlations arise in quantum mechanical systems that mimic the
field theory that we shall consider here, and we refer the reader
to that paper for the role that classical correlations play.  Our
concern in this talk is, rather, with diagonalisation, equally
necessary for the onset of classical behaviour. This is determined
both through the master equation for the evolution of the density
matrix and the decoherence functional, whose role is to describe
consistent histories.

This talk builds upon earlier work by us
\cite{lomplb,lomplb2,lomnpb}, and we refer the reader to it for
much of the technical details. We restrict ourselves to flat
space-time. The extension to non-trivial metrics is
straightforward in principle. See the talk of Lombardo in these
same proceedings \cite{POFer}, which complements this.

\section{Basic Ideas}

 The evolution of a quantum field as it falls out of equilibrium at
a transition is determined in large part by its behaviour at early
times, before interactions have time to take effect. To be
concrete, consider a real scalar field $\phi (x)$, described by a
$Z_2$-symmetry breaking action ($\mu^2 >0$)
\begin{equation}
S[\phi ] = \int d^4x\left\{ \frac{1}{2}\partial_{\mu}
\phi\partial^{\mu} \phi + \frac{1}{2}\mu^2 \phi^2 -
\frac{\lambda}{4!}\phi^4\right\}.
 \label{phi4}
\end{equation}
with symmetry breaking scale $\eta^2 = 6\mu^2/\lambda$. On
heating, this shows a continuous transition, with critical
temperature $T_{\rm c}^2 = 2\eta^2$. If, by virtue of the expansion of
the universe the system is very rapidly cooled (quenched) from
$T > T_{\rm c}$ to $T < T_{\rm c}$, the initial stages of the
transition can be described by a free field theory with inverted mass $-\mu^2<0$.
The state of the field is initially concentrated on the local
maximum of the potential, and spreads out with time. This
description is valid for short times, until the field wave
functional explores the ground states of the potential.

The $\phi$-field ordering after the transition is due to the
growth in amplitude of its {\it unstable} long-wavelength modes,
which we term $\phi_<(x)$. For an instantaneous quench these have
wave-number $k<\mu$ for all time. Although the situation is more
complicated for slower quenches, until the transition is complete
there are always unstable modes. As a complement to these, we
anticipate that the {\it stable} short-wavelength modes of the
field $\phi_>(x)$, where
 \begin{equation}
 \phi (x) = \phi_<(x) + \phi_>(x),
 \nonumber
 \end{equation}
will form an environment whose coarse-graining makes the
long-wavelength modes decohere \cite{fernando}. In practice, the
boundary between stable and unstable is not crucially important,
provided there is time enough for the power in the field
fluctuations to be firmly in the long-wavelength modes. This
requires weak coupling $\lambda\ll 1$.  Of course, all the other
fields with which $\phi$ interacts will contribute to its
decoherence, but for the moment we ignore such fields (before
returning to them in the last section).

After splitting, the action (\ref{phi4}) can be written as
\begin{equation}S[\phi] = S[\phi_<] + S[\phi_>] + S_{\rm int}[\phi_<,
\phi_>],\label{actions}\end{equation} where the interaction term
is dominated \cite{lomplb,lomnpb} by its biquadratic term
\begin{equation}
S_{\rm int}[\phi_<, \phi_>] \approx -\frac{1}{6}\lambda \int d^4x
~ \phi_<^2(x) \phi_> ^2(x).
 \label{inter}
 \end{equation}

The total density matrix (for the system and bath fields) is
defined by
\begin{equation} \rho_{\rm r}[\phi^+,\phi^-,t]=
\rho[\phi_<^+,\phi_>^+,\phi_<^-,\phi_>^-,t]=\langle\phi_<^+
\phi_>^+\vert {\hat\rho} \vert \phi_<^-
\phi_>^-\rangle,\label{matrix} \nonumber
\end{equation}
and we assume that, initially, the thermal system and its
environment are not correlated.

On tracing out the short-wavelength modes, the reduced density
matrix
\begin{equation}\rho_{\rm r}[\phi_<^+,\phi_<^-,t] = \int {\cal D}\phi_>
\rho[\phi_<^+,\phi_>,\phi_<^-,\phi_>,t],\label{red}\nonumber
\end{equation}
whose diagonalisation determines the onset of classical behaviour,
evolves as
\begin{equation}\rho_{\rm r}[t] = \int d\phi_{<i}^+
\int d\phi_{<i}^- ~ J_{\rm r}[t,t_i]~ \rho_{\rm
r}[t_i],\label{evol}\nonumber
\end{equation}
where $J_{\rm r}[t,t_i]$ is the evolution operator
\begin{equation}J_{\rm r}[t,t_i] =
\int_{\phi_{<i}^+}^{\phi_{<f}}{\cal D}\phi_<
\int_{\phi¯_{<i}}^{\phi_{<f}¯}{\cal D}\phi_<¯\,
\exp\{iS_{CG}[\phi_<^+,\phi_<^-]\}.\label{evolred}
\end{equation}
$S_{CG}[\phi_<^+,\phi_<^-]$ is the coarse-grained effective
action, of the closed time-path form
\begin{equation}S_{CG}[\phi_<^+,\phi_<^-] = S[\phi_<^+] - S[\phi_<^-] + \delta
S[\phi_<^+,\phi_<^-].\label{CTPEA}\nonumber\end{equation}
 All the
information about the effect of the environment is encoded in
$\delta S[\phi_<^+,\phi_<^-]$ through the influence functional (or
Feynman-Vernon functional \cite{feynver})
\begin{equation}F[\phi_<^+,\phi_<^-] = \exp \{i
\delta S[\phi_<^+,\phi_<^-]\},\label{IA}\nonumber\end{equation}
giving $\delta S$ a well defined diagrammatic expansion.

\section{The Master Equation}

To see how the diagonalisation of $\rho_{\rm r}$ occurs, we construct the
{\it master equation}, which casts its evolution in differential
form.
 As a first approximation, we make a
saddle-point approximation for $J_r$ in (\ref{evolred}),
\begin{equation}
J_{\rm r} [\phi_{{\rm f}}^+,\phi_{{\rm f}}^-, t_f\vert \phi_{{\rm
i}}^+,\phi_{{\rm i}}^-, t_i] \approx\exp (i
S_{CG}[\phi^+_{\rm cl},\phi^-_{\rm cl}]), \label{saddle}
\end{equation}
In (\ref{saddle}) $\phi^\pm_{\rm cl}$  is the solution to the
equation of motion
\begin{equation}
\frac{\delta Re S_{CG}}{\delta\phi^+}\bigg |_{\phi^+=\phi^-}
=0\nonumber,
\end{equation}
 with boundary conditions $\phi^\pm_{\rm cl}(t_0)=\phi^\pm_{\rm i}$
and $\phi^\pm_{\rm cl}(t)=\phi^\pm_{\rm f}$.

It is very difficult to solve this equation analytically. We
exploit the fact that, after the transition, the field cannot be
homogeneous in one of its groundstates $\phi = \eta$ or $\phi =
-\eta$ because of causality \cite{kibble}. As a result there is an
effective 'domain' structure in which the domain boundaries are
'walls' across which $\phi$ flips from one groundstate to the
other. Further, these domains have a characteristic size $\xi$,
where $\xi^{-1}= \pi k_0$ labels the dominant momentum in the
power of the $\phi$-field fluctuations as the unstable
long-wavelength modes grow exponentially. For simplicity, we adopt a
'minisuperspace' approximation, in which we assume regular
domains, enabling $\phi_{\rm cl}(\vec x, s)$ to be written as
\begin{equation}
\phi_{\rm cl}(\vec x, s) =  f(s,t)\Phi (x)\Phi (y)\Phi (z),
\label{phiclass}\end{equation}
 where
$\Phi (0) = \Phi (\xi) = 0$, and
\[\Phi (x+\xi) = -\Phi (x).
 \label{phicl}
\]
 $f(s,t)$ satisfies $f(0,t)= \phi_{\rm i}$ and $f(t,t) = \phi_{\rm
f}$. We write it as
\begin{equation} f(s,t) = \phi_{\rm i} u_1(s,t) +
\phi_{\rm f} u_2(s,t).
 \label{us}
 \end{equation}
 In \cite{lomnpb} we made the simplest choice for $\Phi(x)$,
 $$ \Phi(x) = \sin k_0x.$$
 Extensions to include more Fourier modes
 are straightforward in principle, but our work in \cite{lomnpb} was sufficient
 to show that the results only depend weakly on the details of the domain function
 $\Phi(x)$ for few Fourier modes.
 In the light of the more
 qualitative comments made here, we refer the reader again to
 \cite{lomnpb} for details.
On the other hand, the $u_i(s,t)$ are solutions of the mode
equation for wavenumber $k_0$ during the quench, with boundary
conditions $u_1(0,t) = 1$, $u_1(t,t) = 0$ and $u_2(0,t) = 0$,
$u_2(t,t) = 1$.

In order to obtain the master equation we must compute the final
time derivative of the propagator $J_{\rm r}$. After that, all the
dependence on the initial field configurations $\phi^\pm_{\rm i}$
(coming from the classical solutions $\phi^\pm_{\rm cl}$) must be
eliminated. Assuming that the unstable growth has implemented
diagonalisation before back-reaction is important, $J_r$ can be
determined, approximately, from the {\it free} propagators as
\begin{equation}J_0[t,t_i]=
\int_{\phi_{<i}^+}^{\phi_{<f}}{\cal D}\phi_<
\int_{\phi¯_{<i}}^{\phi_{<f}¯}{\cal D}\phi_<¯\exp\{i[
S_0(\phi^+) - S_0(\phi^-)]\}\label{propdeJ0}
\end{equation}
where $S_0$ is the free-field action. This  satisfies the general
identities \cite{fernando}
\begin{equation}
\phi^\pm_{\rm cl}(s) J_0 = \Big[\phi^\pm_{\rm f} [u_2(s,t) -
\frac{{\dot u}_2(t,t)}{{\dot
             u}_1(t,t)}u_1(s,t)] \mp  i {u_1(s,t)\over{{\dot u}_1(t,t)}}
\partial_{\phi^\pm_{<\rm f}}\Big]J_0
\label{rel1}\nonumber
\end{equation}
which allow us to remove the initial field configurations
$\phi^\pm_{\rm i}$, and obtain the master equation.

Even with these simplifications the full equation is very
complicated, but it is sufficient to calculate the correction to
the usual unitary evolution coming from the noise (diffusion)
kernels (to be defined later). The result reads
\begin{equation}
i {\dot \rho}_{\rm r} = \langle \phi^+_{<\rm f}\vert [H,\rho_{\rm
r}] \vert \phi^-_{<\rm f}\rangle - i V \Delta^2 D(\omega_0, t)
\rho_{\rm r}+ ... \label{master}
\end{equation}
where $D$ is the diffusion coefficient and $$\Delta = (\phi_{\rm f}^{+2}
- \phi_{\rm f}^{-2})/2$$ for the {\it final} field configurations
(henceforth we drop the suffix). The ellipsis denotes other terms
coming from the time derivative that do not contribute to the
diffusive effects. $V$ is understood as the minimal volume inside
which there are no coherent superpositions of macroscopically
distinguishable states for the field.

The effect of the diffusion coefficient on the decoherence process
can be seen by considering the following approximate solution to
the master equation:
\begin{equation} \rho_{\rm r}[\phi_<^+, \phi_<^-; t]\approx
\rho^{\rm u}_{\rm r}[\phi_<^+, \phi_<^-; t] ~ \exp \left[-V
\Delta^2 \int_0^t ds ~D(k_0, s) \right],\nonumber
\end{equation} where $\rho^{\rm u}_{\rm r}$ is the solution of the
unitary part of the master equation (i.e. without environment).
The system will decohere when the non-diagonal elements of the
reduced density matrix are much smaller than the diagonal ones.

The decoherence time  $t_{D}$ sets the scale after which we have a
classical system-field configuration, and depends strongly on the
properties of the environment. It satisfies
\begin{equation}
1 \approx  V\Delta^2\int_{0}^{t_{\rm D}} ds ~D(k_0,s) ,
\label{Dsum}
\end{equation}
and corresponds to the time after which we are able to distinguish
between two different field amplitudes, inside a given volume $V$.

 To terms up to order $\lambda^2$ and one loop in
the $\hbar$ expansion (we continue to work in units in which
$\hbar = k_B = 1$), the influence action due to the biquadratic
interaction between system and environment has imaginary part
\begin{equation}
{\rm Im}\delta S  = - \int d^4x \int d^4y
 \Delta(x)N(x,y) \Delta
(y),
\end{equation} where
$N(x,y) = \frac{1}{4}\lambda^2 {\rm Re}G^{>2}_{++}(x,y)$ is the
noise (diffusion) kernel and $G^{>}_{++}(x, y)$ is the thermal
short-wavelength closed time-path correlator.

 Explicit calculation shows that $D(k_0,t)$ takes
the form
\begin{equation} D(k_0,t)
= \int_0^t ~ ds ~ u(s,t)~ F(k_0,s,t) \label{D0}
\end{equation}
where
$$u(s,t) =\bigg[u_2(s,t) - \frac{{\dot u}_2(t,t)}{{\dot
              u}_1(t,t)}u_1(s,t)\bigg]^{2},$$
and $ F(k_0,s,t)$ is built from the spatial Fourier transforms of
the overlap of the diffusion kernel with the field profiles $\Phi
(x)\Phi (y)\Phi (z)$.

In the integrand of (\ref{D0}) $u(s,t)$ is rapidly varying, driven
by the unstable modes, and $F(k_0,s,t)$ is slowly varying. For
long-wavelengths $k_0\ll\mu$ we have, approximately,
$$ F(k_0,s,t) = O(N(k_0 =0; t-s)),$$ whereby
\begin{equation}
 D(k_0,t) \approx   F(k_0, 0,t) ~
\int_0^{t} ds ~ u(s,t). \label{D02}
\end{equation}
That is, the diffusion coefficient factorises into the
environmental term $F$, relatively insensitive to both wavenumber
and time, and the rapidly growing integral that measures the
classical growth of the unstable system modes that are ordered in
the transition.

To be specific, we restrict ourselves to the simplest case of an
instantaneous quench from a temperature $T = {\cal O}(T_c) > T_C$,
for which
 \begin{equation}
 u_1=  {\sinh[\omega_0 (t -
s)] \over{\sinh(\omega_0 t)}},\,\,u_2(s,t)=  {\sinh(\omega_0 s)
\over {\sinh(\omega_0 t),}},\,\,\, \label{us2}
\end{equation}
where $\omega_0^2 = \mu^2 - k_0^2\approx \mu^2$. It follows that
\begin{equation}
u(s,t) = \cosh^2[\omega_0(t-s)],
\end{equation}
from whose end-point behaviour at $s=0$ of the integral
(\ref{D02}) we find the even simpler result
\begin{equation}
D(k_0,t)\sim \mu^{-1}F(k_0, 0,t)~ u(0,t)
\sim (\lambda T_{\rm c}/4\pi\mu)^2~ \exp [2\mu t],
\label{D(t)}
\end{equation}
assuming $\mu t_D\gg 1$.

For more general quenches growth is more complicated than simple
exponential behaviour but a similar separation into fast and slow
components applies.

We have omitted a large amount of complicated technical detail
(see \cite{lomnpb}), to give such a simple final result. This
suggests that we could have reached the same conclusion more
directly.

We now indicate how we can obtain the same results by demanding
consistent histories of the $\phi$ field.

\section{The Decoherence Functional}

The notion of consistent histories provides a parallel approach to
classicality. Quantum evolution can be considered as a coherent
superposition of fine-grained histories. If one defines the
c-number field $\phi (x)$ as specifying a fine-grained history,
the quantum amplitude for that history is $\Psi [\phi] \sim
e^{iS[\phi]}$ (we continue to work in units in which $\hbar =1$).

In the quantum open system approach that we have adopted here, we
are concerned with coarse-grained histories
\begin{equation}
\Psi [\alpha] = \int {\cal D}\phi ~ e^{iS[\phi]}\alpha [\phi]
\end{equation}
where $\alpha [\phi]$ is the filter function that defines the
coarse-graining.

From this we define the decoherence function for two
coarse-grained histories as
\begin{equation}
 {\cal D}[\alpha^+,\alpha^-] = \int {\cal D}\phi^+{\cal
 D}\phi^-~e^{i(S[\phi^+]-S[\phi^-])}\alpha^+ [\phi^+]\alpha^-
 [\phi^-].
\end{equation}
${\cal D}[\alpha^+,\alpha^-]$ does not factorise because the
histories $\phi^{\pm}$ are not independent; they must assume
identical values on a spacelike surface in the far future.

A necessary and sufficient condition for the validity of the sum
rules of probability theory (i.e. no quantum interference terms)
is \cite{Gri}
\begin{equation}
 {\rm Re}{\cal D}~[\alpha^+,\alpha^-]\approx 0,
\end{equation}
when $\alpha^+\neq\alpha^-$ (although in most cases the stronger
condition ${\cal D}[\alpha^+,\alpha^-]\approx 0$ holds
\cite{Omn}). Such histories are consistent \cite{GH}.

For our particular application, we wish to consider as a single
coarse-grained history all those fine-grained ones where the full
field $\phi$ remains close to a prescribed classical field
configuration $\phi_{\rm cl}$. The filter function takes the form
\begin{equation}
 \alpha_{\rm cl}[\phi ] = \int {\cal D}J~
 e^{i\int J(\phi - \phi_{\rm cl})}\alpha_{\rm cl}[J].
\end{equation}
In principle, we can examine general classical solutions for their
consistency but, in practice, it is simplest to restrict ourselves
to solutions of the form (\ref{phiclass}). In that case, we have made
a de facto separation into long and short-wavelength modes
whereby, in a saddle-point approximation over $J$,
\begin{equation}
 {\cal D}(\phi^+_{\rm cl},\phi^-_{\rm cl}) \sim
 \exp \{iS_{CG}[\phi^+_{\rm cl},\phi^-_{\rm cl}]\}.
\end{equation}
As a result,
 \begin{equation}
|{\cal D}(\phi^+_{\rm cl},\phi^-_{\rm cl})| \sim
 \exp \{-{\rm Im}\delta S[\phi^+_{\rm cl},\phi^-_{\rm cl}]\}
 \end{equation}
For the instantaneous quench of (\ref{us}), using the late time
behaviour $\phi_{\rm cl}^{\pm}\sim e^{\mu s}\phi_0^{\pm}$, ${\rm
Im}\delta S[\phi^+_{\rm cl},\phi^-_{\rm cl}]$ takes the form
\begin{equation}
{\rm Im}~\delta S \sim \frac{V\Delta^2}{\mu^2}\int_0^t ds\int_0^t
ds' e^{2\mu s}~e^{2\mu s'} F(k_0,s,s'). \label{ImS}
\end{equation}

From this viewpoint adjacent histories become consistent at the
time $t_D$, for which
\begin{equation}
 1\approx \int_0^{t_D} dt ~{\rm Im~\delta S}.
\label{tD2}
\end{equation}

\section{The decoherence time}

We have used the same terminology for the time $t_D$ since, on
inspection, (\ref{tD2}) is {\it identical} to (\ref{Dsum}) in
defining the onset of classical behaviour. As we noted, in
practice the use of the decoherence functional looks to be  less
restrictive than the master equation, and we hope to show this
elsewhere.

For the moment what is of interest is whether $t_D$, based on
linearisation of the model, occurs before backreaction sets in, to
invalidate this assumption. When all the details are taken onto
account, whether from (\ref{us}) or (\ref{us2}), $t_D$ satisfies
\begin{equation}
1= {\cal
O}\bigg(\frac{\lambda^2VT_c^2}{\mu^3}\Delta^2\bigg)~\exp(4\mu t_D)
 \label{tD3}
\end{equation}
or, equivalently
\begin{equation}
\exp(4\mu t_D) = {\cal
O}\bigg(\frac{\mu^3}{\lambda^2VT_c^2\Delta^2}\bigg)
\end{equation}
For the rapid quenches considered here, linearisation manifestly
breaks down by the time $t^*$, for which $\langle
\phi^2\rangle_{t^*} \sim \eta^2$, given by
\begin{equation}
\exp (2\mu t^*) =  {\cal O}\bigg( \frac{\mu}{\lambda T_{\rm
c}}\bigg).
 \label{tstar}
\end{equation}
The exponential factor, as always, arises from the growth of the
unstable long-wavelength modes. The factor $T_{\rm c}^{-1}$ comes from
the $\coth(\beta\omega /2)$ factor that encodes the initial
Boltzmann distribution at temperature $T\gtrsim T_{\rm c}$.

Our conservative choice is that the volume factor $V$ is ${\cal
O}(\mu^{-3})$  since $\mu^{-1}$ (the Compton wavelength) is the
smallest scale at which we need to look. With this choice it
follows that
\begin{equation}
\exp 2(t^* - t_D) = {\cal O}\bigg(\frac{|\Delta|}{\mu^2}\bigg)) =
{\cal O}(\bar\phi\delta), \label{tstartD}
\end{equation}
where $\bar\phi = (\phi_{<}^+ + \phi_{<}^-)/2\mu,$ and $ \delta =
|\phi_{<}^+ - \phi_{<}^-|/2\mu$. Within the volume $V$ we do not
discriminate between field amplitudes which differ by $ {\cal
O}(\mu) $, and therefore take $\delta = {\cal O}(1)$. For
$\bar\phi$ we note that, if $t_D$ were to equal $t^*$, then
$\bar\phi^2 = {\cal O}(1/\lambda)\gg 1,$ and in general $\bar\phi
>1$. As a result, if there are no large numerical factors, we have
\begin{equation}
t_D < t^*, \label{tllt}
\end{equation}
and the density matrix has become diagonal before the transition
is complete.  Detailed calculation shows \cite{lomnpb} that there
are no large factors \cite{footnote}.

We already see a significant difference between the behaviour for
the case of a biquadratic interaction  with an environment given
in (\ref{Dsum}) and the more familiar linear interaction usually
adopted in quantum mechanics. This latter would have replaced
$\Delta/\mu^2$ just by $\delta$,  incapable of inducing
decoherence before the transition is complete.

We note that, once the interaction strength is sufficiently weak
for classical behaviour to appear before the transition is
complete, this persists, however weak the coupling becomes. It
remains the case that, the  weaker the coupling, the longer it
takes for the environment to decohere the system but, at the same
time, the longer it takes for the transition to be completed, and
the ordering (\ref{tllt}) remains the same. This is equally true
for more general quenches provided the system remains
approximately Gaussian until the transition is complete.

\section{Extensions of the model}

Finally, it has to be said that taking only the short wavelength
modes of the field as a one-loop system environment is not a
robust approximation. We should sum over hard thermal loops in the
$\phi$-propagators. To be in proper control of the diffusion we
need an environment that interacts with the system, without the
system having a strong impact on the environment. We are helped in
that, in the early universe, the order parameter field $\phi$ will
interact with any field $\chi$ for which there is no selection
rule. Again, it is the biquadratic interactions that are the most
important.

The most simple additional environment is one of a large number
$N\gg 1$ of weakly coupled scalar fields $\chi_{\rm a}$, for which
the action (\ref{phi4}) is extended to

\begin{equation}
S[\phi , \chi ] = S[\phi ] + S[\chi ] + S_{\rm int}[\phi ,\chi ],
\label{quaction}
\end{equation}
where $S[\phi]$ is as before, and
\begin{eqnarray}
&&S[\chi_{\rm a} ] = \sum_{\rm a=1}^N\int d^4x\left\{
{1\over{2}}\partial_{\mu}\chi_{\rm a}
\partial^{\mu}
\chi_{\rm a} - {1\over{2}} m_{\rm a}^2 \chi^2_{\rm a}\right\}, \nonumber
\\
&&S_{\rm int}[\phi ,\chi ] = - \sum_{\rm a=1}^N\frac{g_{\rm a}}{8}
\int d^4x \phi^2 (x) \chi^2_{\rm a} (x), \label{Sint}
\end{eqnarray}
where $m_{\rm a}^2 >0$. For simplicity we take weak couplings $\lambda
\simeq g_{\rm a}$ and comparable masses $m_{\rm a}\simeq \mu$. The
effect of a large number of weakly interacting environmental
fields is twofold. Firstly, the $\chi_a$ fields reduce the
critical temperature $T_{\rm c}$ and, in order that $T_{\rm c}^2=\frac
{2\mu^2}{\lambda + \sum g_{\rm a}}\gg \mu^2$, we must take
$\lambda + \sum g_{\rm a} \ll 1$.  Secondly, the single
$\chi$-loop contribution to the diffusion coefficient is the
dominant $\chi$-field effect if, for order of magnitude estimates,
we take identical $g_{\rm a} =\bar g/\sqrt{N}$, whereby $1\gg
1/\sqrt N\gg\bar g\simeq \lambda$. With this choice the effect of
the $\phi$-field on the $\chi_{\rm a}$ thermal masses is, relatively,
$O(1/\sqrt{N})$ and can be ignored. We stress that this is not a
Hartree or large-N approximation of the type that, to date, has
been the main way to proceed \cite{boya,mottola,Greg} for a {\it
closed} system.

Provided the change in temperature is not too slow the exponential
instabilities of the $\phi$-field grow so fast that the field has
populated the degenerate vacua well before the temperature has
dropped to zero. Since the temperature $T_c$ has no particular
significance for the environment field, for these early times we
can keep the temperature of the environment fixed at $T_{\chi}
={\cal O}(T_{\rm c})$ (our calculations are only at the level of
orders of magnitude). As before, we split the field as $\phi =
\phi_< + \phi_>$. The $\chi$-fields give an additional one-loop
contribution to $D(k_0, t)$ with the same $u(s)$ but a $G_{++}$
constructed from (all the modes of) the $\chi$-field. The
separation of the diffusion coefficient due to $\chi$ into fast
and slow factors proceeds as before to give a term that is
identical to (\ref{D(t)}) (or (\ref{ImS})) but for its ${\bar
g}^2$ prefactor.

Diffusion effects are {\it additive} at the one-loop level, and
the final effect is to replace $\lambda^2$ in (\ref{tD3}) by
$\lambda^2 + {\bar g}^2 >\lambda^2$, while leaving (\ref{tstar})
unchanged. Although the relationship between $T_{\rm c}$ and $\lambda$
has been uncoupled by the presence of the $\chi_{\rm a}$, the
relationship (\ref{tstartD}) persists, with an enhanced right hand
side, requiring that (\ref{tllt}) is even better satisfied.

Given that the effect of further environmental fields is to
increase the diffusion coefficient and speed up the onset of
classical behaviour, additional fields interacting with the $\phi$
field seem superfluous. However, the symmetries of the universe
seem to be local (gauge symmetries), rather than global, and we
should take gauge fields into account. We conclude with some
observations from our work in progress \cite{ubaIC} with local
symmetry breaking.

Local symmetry breaking is not possible for our real $\phi$ field
but, as a first step \cite{lomplb2}, it is not difficult  to
extend our model to that of a complex $\phi$-field. At the level
of $O(2)$ global interactions with external fields and with its
own short-wavelength modes, things are largely as before. Local
$U(1)$ symmetry breaking is most easily accommodated by taking the
$\phi$-field to interact with other charged fields $\chi$ through
the local $U(1)$ action
\begin{equation}
S [\phi, A_{\mu},\chi ] = S [\phi, A_{\mu}] + S_{\chi}
[A_{\mu},\chi ],
 \label{S}
\end{equation}
in which $S[\phi, A_{\mu} ] =$
\begin{equation}
\int d^4x\left\{(D_{\mu} \phi)^*D^{\mu} \phi + \mu^2 \phi^*\phi -
{\lambda\over{4}}(\phi^*\phi)^2
-\frac{1}{4}F^{\mu\nu}F_{\mu\nu}\right\}, \label{SQED}
\end{equation}
and
\begin{equation}
S [A_{\mu}, \chi ] = \int d^4x\left\{(D_{\mu} \chi)^*D^{\mu} \chi
+ m^2\chi^*\chi\right\}. \label{chi}
\end{equation}
For simplicity we have taken a single $\chi$ field. The theory
(\ref{S}) shows a phase transition at temperature $T_c$, and we
assume couplings are such as to make this transition continuous.
At the level of one loop the additional term to the diffusion
function has derivative couplings. Having made a gauge choice,
these give rise to explicit momenta factors $k_{\mu}$ in the
generalisation of $F$. Unlike the contributions to $D$ that we
have seen so far, which are largely insensitive to the momentum
scale $k_0$, these contributions are strongly damped at large
wavelength. In consequence, they barely enhance the onset of
classical behaviour but, given that the effect of the other
environmental modes is to enforce classical behaviour so quickly,
it hardly matters.

\section{Conclusion}

The previous paragraph says it all. For fast quenches weakly
coupled environments make a scalar order parameter field decohere
before the transition is complete, under very general assumptions.
An essential ingredient for rapid decoherence is {\it nonlinear}
coupling to the environment, inevitable when that environment
contains the short-wavelength modes of the order parameter field.
Had we only considered linear coupling to the environment, as in
\cite{kim}, for example (but an assumption that is ubiquitous in
quantum mechanical models, from Brownian motion onwards)
decoherence would not have happened before the transition was
complete, and we would not know how to proceed, although classical
correlations would have occurred.

\section{acknowledgments}
 We thank Diego Mazzitelli for his
collaboration in this work. F.C.L. was supported by Universidad de
Buenos Aires, CONICET (Argentina), Fundaci\'on Antorchas and
ANPCyT. R.J.R. was supported in part by the COSLAB programme of
the European Science Foundation.


\begin{references}
\bibitem{footnote} In evaluating $F$ we perform the
loop diagram using the full propagator rather than just its short
wavelength modes. Since the $T_{\rm c}$ behaviour comes entirely from
the short wavelength part of the integral, this is justified
\end{references}
\end{document}